\documentclass[12pt]{article}

\pdfoutput=1

\usepackage[top=80pt,bottom=80pt,left=65pt,right=65pt]{geometry}
\usepackage{amssymb,amsmath,xypic}
\usepackage{graphicx,color,subfigure}
\usepackage{float}
\usepackage{sectsty}
\usepackage{cite}
\usepackage[debug,pageanchor=false]{hyperref}
\usepackage{slashed}
\definecolor{link}{rgb}{.8,.15,.1}
\hypersetup{colorlinks=true,linkcolor=link,citecolor=link,urlcolor=link,linktocpage}

\makeatletter
\@addtoreset{equation}{section}
\makeatother

\newcommand{\beq}{\begin{equation}}
\newcommand{\eeq}{\end{equation}}
\newcommand{\bea}{\begin{eqnarray}}
\newcommand{\eea}{\end{eqnarray}}
\newcommand{\nn}{\nonumber}

\begin{document}

\begin{titlepage}

\begin{center}

\vskip .5in 
\noindent

{\Large \bf{Type II Solutions on AdS$_3\times S^3\times S^3$ with Large Superconformal Symmetry}}

\bigskip\medskip

 Niall T. Macpherson\\

\bigskip\medskip
{\small

 SISSA International School for Advanced Studies\\
Via Bonomea 265, 34136 Trieste \\
			and\\ INFN, sezione di Trieste
	
}

\vskip .5cm 
{\small \tt nmacpher@sissa.it}
\vskip .9cm 
     	{\bf Abstract }

\vskip .1in
\end{center}

\noindent
New local solutions in type II supergravity that are foliations of  AdS$_3\times S^3\times S^3$ over an interval and preserve at least large $\mathcal{N}=(4,0)$ supersymmetry are found. Some cases have compact internal space, some not and one experiences an enhancement to $\mathcal{N}=(4,4)$. We present two new globally compact solutions with D brane and O plane sources explicitly, one in each of IIA and IIB. The former is part of an infinite family of solutions with D8/O8s back reacted on AdS$_3\times S^3\times S^3\times S^1$. In the latter the the algebra degenerates to small $\mathcal{N}=(4,0)$ and the internal geometry is bounded between D5s and O5s back reacted on AdS$_3\times S^3\times \mathbb{R}^4$.

\noindent
 
\vfill
\eject

\end{titlepage}

\tableofcontents

\section{Introduction}
The AdS-CFT correspondence has by now shown itself to be a powerful tool to probe the dynamics of theories on both sides of the correspondence. Since its inception it has stimulated progress constructing many  CFT$_d$s and their dual AdS$_{d+1}$ solutions, in many cases embedded into 10 dimensions. One area where progress on the CFT side has somewhat outpaced the other is the  AdS$_3$-CFT$_2$ correspondence. This is not to say that progress on the gravity side has not been made (see \cite{Boonstra:1998yu,Giveon:1998ns,Elitzur:1998mm,deBoer:1999gea,Argurio:2000tg,Gukov:2004ym,Kim:2005ez,Gauntlett:2006af,Donos:2008hd,DHoker:2008lup,Estes:2012vm,Bachas:2013vza,Jeong:2014iva,Kelekci:2014ima,Donos:2014eua,Lozano:2015bra,Karndumri:2015sia,Kelekci:2016uqv,Eberhardt:2017fsi,Couzens:2017way,Couzens:2017nnr,Eberhardt:2017uup,Datta:2017ert,Eberhardt:2018sce} for an incomplete list), merely that there is much more yet to be studied.

Two dimensional CFTs play an important role in physics, in string theory and beyond so there is clear motivation to construct holographic duals. The barrier to this is that when embedded into 10 dimensional supergravity, their internal space is 7-dimensional which is rather large. Progress can be made tractable by assuming extended supersymmetry - in this case the dual geometry will realise an additional R-symmetry reducing the number of undetermined dimensions. An interesting feature of superconformal field theories in 2 dimensions is that a relatively large number of superconformal algebras exist for each  number of preserved supercharges,  with each preserving a distinct R-symmetry. Those that can be embedded into 10 and 11 dimensions were classified in \cite{Beck:2017wpm}. Given this, and the recent G-structure classification of $\mathcal{N}=1$ AdS$_3$ solutions in type II supergravity \cite{Dibitetto:2018ftj}, the time seems right to begin to seriously explore the possibilities.

An interesting class of AdS$_3$ solutions with limited examples exhibiting a compact internal space (as required for a holographic dual of a 2d CFT) are those preserving at least $\mathcal{N}=(4,0)$ supersymmetry with the so-called ``large'' superconformal algebra $\mathfrak{d}(2,1,\alpha)$. This has maximal bosonic sub-algebra $\mathfrak{s}\mathfrak{l}(2)\oplus \mathfrak{so}(4)$, where the second term is an SO(4) R-symmetry. Large $\mathcal{N}=(4,0)$ includes a Kac-Moody algebra  $\mathfrak{su}(2)_{k_+}\oplus\mathfrak{su}(2)_{k_-}$ (in contrast to small $\mathcal{N}=(4,0)$ which has just $\mathfrak{su}(2)_{k}$) and CFTs with this symmetry are characterised by the following relations between their central charge $c$, the continuous parameter $\alpha$, and the levels $k_{\pm}$ \cite{Sevrin:1988ew} 
\beq\label{eq:chargealpha}
c= 6\frac{ k_+ k_-}{k_++ k_-}.~~~~~ \alpha = \frac{k_-}{k_+}.
\eeq
Precisely what is assigned to be $k_{+}$ and $k_{-}$ appears ambiguous in $c$, but this is a manifestation of the fact that $\mathfrak{d}(2,1,\alpha)$ is isomorphic to $\mathfrak{d}(2,1,\alpha^{-1})$ -  as such the central charge is sufficient to determine the value of $\alpha$ (up to identifying $\alpha \sim \alpha^{-1}$) for a given large $\mathcal{N}=(4,0)$ CFT$_2$.
 
The canonical example of a supergravity solution dual to a CFT$_2$ with large superconformal symmetry is AdS$_3\times S^3\times S^3\times S^1$ (see also \cite{Elitzur:1998mm} for a worldsheet perspective) and its M-theory avatar AdS$_3\times S^3\times S^3\times T^2$ \cite{Boonstra:1998yu,deBoer:1999gea}  which actually preserve large $\mathcal{N}=(4,4)$ supersymmetry (a maximal case for AdS$_3$ \cite{Haupt:2018gap}) with algebra $\mathfrak{d}(2,1,\alpha)\oplus \mathfrak{d}(2,1,\alpha)$, and where $\alpha$ is related to the radii of the 3-spheres. In M-theory such solutions were classified locally then globally in \cite{DHoker:2008lup,Estes:2012vm,Bachas:2013vza} where those consistent with a dual CFT$_2$ were claimed to be locally AdS$_3\times S^3\times S^3\times T^2$ and so all reduce to AdS$_3\times S^3\times S^3\times S^1$ in IIA (at least locally). Historically there was some difficulty ascertaining the CFT dual to AdS$_3\times S^3\times S^3\times S^1$ \cite{Gukov:2004ym}, though a recent attempt was made in \cite{Tong:2014yna}. In large part this difficulty was due to the failure of otherwise likely CFT proposals to reproduce the BPS spectrum of the supergravity solution calculated in \cite{deBoer:1999gea}. However this computation was recently found to be in error \cite{Eberhardt:2017fsi} and the corrected spectrum was explicitly shown to match that of a certain symmetric orbifold ($\mathcal{S}_k$ \cite{Gukov:2004ym}) in \cite{Eberhardt:2017pty}\footnote{Actually the match between $\mathcal{S}_k$ and AdS$_3\times S^3\times S^3\times S^1$ additionally requires that some flux charges are tuned.}. 

Beyond the cases with maximal supersymmetry (for AdS$_3$), solutions with $\mathcal{N}=(4,0)$ were constructed from AdS$_3\times S^3\times S^3\times S^1$ using T-duality (and its non-abelian counter part) in \cite{Kelekci:2014ima,Lozano:2015bra} and a class of AdS$_3\times S^2\times S^2\times \text{CY}_2$ solutions in M-theory was found in \cite{Kelekci:2016uqv}. Another interesting example is a flow from AdS$_5\times T^{1,1}$  to a twice T-dualised version of AdS$_3\times S^3\times S^3\times S^1$ preserving $\mathcal{N}=(4,2)$ \cite{Donos:2014eua} (other flows across dimensions were found in  \cite{DHoker:2008lup,Estes:2012vm,Bachas:2013vza}, but these exhibit large $\mathcal{N}=(4,4)$). Finally, somewhat related to this story, there is also a family of $\mathcal{N}=(2,0)$ solution in IIB that are AdS$_3\times S^3\times S^3\times S^1$ only topologically \cite{Donos:2008hd}.
\\
\\
In this work new AdS$_3$ preserving large $\mathcal{N}=(4,0)$ supersymmetry will be constructed that are neither locally AdS$_3\times S^3\times S^3\times S^1$ nor related to it by duality. To do this one needs to arrange for the internal space to realise an SO(4) R-symmetry.  There are several ways to arrange for this to happen with products of 2 and 3-spheres. Here it will be assumed that the R-symmetry is realised with a foliation of $S^3\times S^3$ over an interval\footnote{Solutions with a similar local foliation were recently constructed in \cite{Dibitetto:2018iar,Dibitetto:2018gtk} by utilising Romans $F_{(4)}$ gauged supergravity.}. Generically, such solutions will have a flavour SO(4) in addition to the R-symmetry\footnote{We shall impose that this entire SO(4)$\times$SO(4) is preserved by the remaining physical fields also.}. The reasons to make this choice are two fold: i) In short, it is the easiest example to look at. However this simplicity will allow for a complete local description of all such solution in type II supergravity\footnote{The same methods could also be used to probe the space of M theory solutions.}. Additionally this should aid the process of finding a CFT dual. ii) With  $S^3\times S^3$ there is the possibility of an enhancement to large $\mathcal{N}=(4,4)$ supersymmetry, thereby generalising the classification of \cite{DHoker:2008lup,Estes:2012vm,Bachas:2013vza} to type II supergravity. One should appreciate though that the assumption of $S^3\times S^3$ limits the scope of this work to a small portion of the space of possible solutions with large $\mathcal{N}=(4,0)$. It will turn out that this portion is far from empty, but one should view this as a first step in a much broader classification endeavour with most solutions lying outside this ansatz.

The method used here to find new solutions with large $\mathcal{N}=(4,0)$ supersymmetry shall be to construct spinors  manifestly realising the bosonic sub-algebra of $\mathfrak{d}(2,1,\alpha)$
\beq
\mathfrak{s}\mathfrak{l}(2)\oplus \mathfrak{so}(4)\nn.
\eeq 
The first factor will be realised by Killing spinors on AdS$_3$ - it  requires a little more work to construct general spinors on the internal space that manifestly transform under the action of SO(4). Having such spinors we shall then find every solution with an $S^3\times S^3$ factor consistent with them. This follows the line of reasoning of the earlier works \cite{Macpherson:2016xwk,Macpherson:2017mvu,Apruzzi:2018cvq,DeLuca:2018buk,Legramandi:2018itv}, where many of the technical details exploited here were originally worked out. Here it will be possible to give the explicit local form of every type II solution consistent with the SO(4) spinor. Strictly speaking, as one is not imposing the entire superconformal alebra $\mathfrak{d}(2,1,\alpha)$, the solutions that follow could in fact preserve some other algebra with bosonic sector containing $\mathfrak{s}\mathfrak{l}(2)\oplus \mathfrak{so}(4)$. One possibility is that SO(4) lies within a larger R-symmetry but given the ansatz for the internal space, the only possibilities for enhanced R-symmetries (other than SO(4)$\times$ SO(4)) are SO(8) and Spin(7) \cite{Beck:2017wpm} which require the internal space to become $S^7$ \cite{Dibitetto:2018ftj}. However $S^7$ is never realised by the solutions constructed here. The other possibility is that the SO(4) R-symmetry of the geometry is realising an SU(2) R-symmetry of the dual CFT and an additional SU(2) outer automorphism symmetry, as is the case with AdS$_2\times S^3\times \mathbb{R}^4$. Such solutions are  degenerate cases of $\mathfrak{d}(2,1,\alpha)$  with $\alpha\to 0$ where small $\mathcal{N}=(4,0)$ is recover - we will find one such example in our analysis. That leaves the question of how one calculates $\alpha$ - one could proceed as in \cite{DHoker:2008wvd} and carefully map bi-linears of the spinors on AdS$_3\times S^3\times S^3$ to the algebra and compute $\alpha$ directly. However, for the examples with compact internal space (the only ones dual to well defined 2d CFTs), there is an easier way. One simply computes the holographic central charge, and then read $\alpha$ off from \eqref{eq:chargealpha} - this will be the route followed here.  
\\
\\
The outline of the paper is as follows: In section \ref{sec:one} we explicitly construct general spinors that transform in the fundamental representation of one of the two available independent SO(4) isometries on $S^3\times S^3$, that are also singlets under the action of the other - this ensures we are consistent with $\mathcal{N}=(4,0)$ supersymmetry and an SO(4) R-symmetry. In section \ref{sec:two} we use G-structure techniques to extract geometric conditions from the SO(4) spinors that all solutions should obey, and in sections \ref{sec:IIA} and \ref{sec:IIB} we find all local solutions that follow. The most interesting of these are clearly those that can be used to construct global solutions with compact internal space. We explicitly construct two such examples (though in IIA infinitely many are possible): A new $\mathcal{N}=(4,0)$ massive IIA solution with large superconformal symmetry in section \ref{eq:case1IIA} and a new $\mathcal{N}=(4,0)$ IIB solution in section \ref{eq:case1IIB} with small superconformal symmetry and SU(2) outer automorphism symmetry. The IIA solution is constructed by gluing two locally non compact solutions together with a D8 brane defect. Further new local solutions preserving $\mathcal{N}=(4,4)$ and $\mathcal{N}=(4,0)$ can be found in sections \ref{eq:case2IIA} and \ref{eq:case2IIB} respectively, but these are neither compact nor related to flows across dimensions (at not least obviously). In section \ref{eq:case2IIA} we speculate that these might also be used to construct globally compact solutions by using (this time) smeared Dp or NS5 brane defects for $p<8$ to glue them together, however a detailed study is beyond the scope here. Finally appendix \ref{sec:conventions} details the conventions used throughout and \ref{sec:proof} proves a claim made in \ref{sec:one}.

Given the results here  and  \cite{Dibitetto:2018ftj}, where AdS$_3$ solutions with exceptional R-symmetries were studied,  it has become clear that such R-symmetry based spinor constructions are a powerful tool to study AdS$_3$ solutions with extended supersymmetry.
\section{Realising an SO(4) R-symmetry on $S^3\times S^3$}\label{sec:one}
In this section we will construct Killing spinors that realise the bosonic sub-algebra of $\mathfrak{d}(2,1,\alpha)$, the conventions used and explicit representations of the of what follows can be found in appendix \ref{sec:conventions}.

We are interested in large $\mathcal{N}=(4,0)$ AdS$_3$ solutions preserving an SO(4) R-symmetry. Any  AdS$_3$ solution can  be expressed in the form
\begin{align}\label{eq:ads3sol}
ds^2&= e^{2A} ds^2(\text{AdS}_3)+ ds^2(\text{M}_7),\nn\\[2mm]
F&= f+ e^{3A}\text{Vol}(\text{AdS}_3)\wedge\star_7\lambda( f),\quad H= h_0\text{Vol}(\text{AdS}_3)+ H_3,
\end{align}
where $F$ is the RR polyform\footnote{In IIA it is $F=F_0+ F_2+F_4+ F_6+ F_8 + F_{10}$, in IIB it is $F=F_1+ F_3+F_5+ F_7+ F_9$.} and $H$ the NS 3-form fluxes. Both of these decompose in terms of their magnetic components $f$, $H_3$ which are defined on $M_7$ only and electric counterparts with legs on all $\text{AdS}_3$ directions ensuring that fluxes respect the isometry of $\text{AdS}_3$. The Bianchi identity $dH=0$ fixes $h_0$ to be constant, while the electric component of $F$ is fixed such the lower and higher fluxes are correctly related under 10d hodge duality - for this reason the operator $\lambda$ is defined such that
\beq\label{eq:lambdadef}
\lambda(X_n)= (-)^{\frac{n}{2}(n-1)} X_n
\eeq
 when acting on an $n$-form. Finally the AdS warp factor $e^{2A}$, and likewise the dilaton $\Phi$ have support on M$_7$ only and the RR polyform obeys
\beq
dF-H\wedge F=0,
\eeq
away from localised sources.

As we seek solutions preserving $\mathcal{N}=(4,0)$ the 10 dimensional Majorana spinors may be written as
\beq\label{eq10dspinors}
\epsilon_1=\sum_{I=1}^4 \zeta^I \otimes v_+\otimes \chi^I_1,~~~~\epsilon_2= \sum_{I=1}^4\zeta^I \otimes v_{\mp}\otimes \chi^I_2 
\eeq
where $\zeta^I$ and $\chi^I_{1,2}$ are 4 independent spinors on AdS$_3$ and M$_7$ respectively and $v_{\pm}$ is an auxiliary 2 vector, that is always required when decomposing an even dimensional spinor in terms of 2 odd ones - $\pm$ refers to chirality, so the upper/lower signs are taken in IIA/B. The spinors on AdS$_3$ are Killing, so they obey the equation 
\beq\label{eq:ads3spinor}
\nabla^{\text{AdS}_3}_{a}\zeta^I= \frac{\mu}{2} \gamma^{\text{AdS}_3}_{a}\zeta^I
\eeq
where  $\frac{\mu}{|\mu|}=\pm 1$ parametrise a spinor that is charged under the SL(2)$_{\text{L}/\text{R}}$ subgroup of\\ SO(2,2)$\cong $SL(2)$_{\text{L}}\times $SL(2)$_{\text{R}}$  and is a singlet under SL(2)$_{\text{R}/\text{L}}$. 

As we want an SO(4) R-symmetry, $\chi^I_{1,2}$ should transform in fundamental of this group and solutions should admit a local description with SO(4) realised geometrically. There are several ways to do this, but from the perspective of finding solutions, the simplest way to realise this R-symmetry is to decompose the internal space as a foliation of $S^3_1\times S^3_2$ over an interval in which the  physical fields have support only, i.e. we take the internal metric to be
\beq\label{eq:M7}
ds^2(\text{M}_7)= e^{2k}dr^2+ e^{2C_1}ds^2(S^3_1)+ e^{2C_2}ds^2(S^3_2)
\eeq
where the functions $e^{2k}, e^{2C_1}, e^{2C_2}$ and also now  $e^{A}$ and $\Phi$ depend on $r$ only and we impose that the fluxes depend on $S^3_1$ and $S^3_2$ through their respective volume forms only. This gives us an SO(4)$_1\times$SO(4)$_2$ isometry on M$_7$ to work with allowing for enhancements to $\mathcal{N}=(4,4)$ supersymmetry with SO(4)$\times$SO(4) R-symmetry whenever the physical fields obey certain constraints we shall discuss at the end of the section. A Killing spinors on $S_{1,2}^3$, $\xi$, obeys the equations
\beq\label{eq:S3KSE}
\nabla^{S^3}_i \xi=\frac{i \nu}{2}\gamma^{S^3}_i\xi,~~~~~\nabla^{S^3}_i \xi^c=\frac{i \nu}{2}\gamma^{S^3}_i\xi^c
\eeq
for $\xi^c$ the Majorana conjugate of $\xi$. This time  $\frac{\nu}{|\nu|}=\pm 1$  parametrise a spinor charged under the SU(2)$_{\text{L}/\text{R}}$ subgroup of SO(4) $\cong$ SU(2)$_\text{L}\times$SU(2)$_{\text{R}}$ that are singlets under the SU(2)$_{\text{R}/\text{L}}$. As shown in \cite{Macpherson:2017mvu}, in the Hopf fibration frame of $S^3$ \eqref{eq:hopfframe}, the doublets of SU(2)$_{\text{L}/\text{R}}$ are simply
\beq\label{eq:spinordoublet}
\xi^a = \left(\begin{array}{c}\xi\\ \xi^c\end{array}\right)^a,
\eeq
and obey the $S^3$ Killing spinor equation \eqref{eq:S3KSE} component by component.
On $S^3$ there are two sets of one-forms that are charged under SU(2)$_{\text{L}}$ and SU(2)$_{\text{R}}$ that are dual to the corresponding  SU(2)$_{\text{L}/\text{R}}$ Killing vectors. We parametrise these in a unified language as $K_i$ such that
\beq\label{eq:doneform}
dK_i+ \frac{\nu}{2}\epsilon_{ijk} K_j\wedge K_k=0,
\eeq
where the sign of $\nu$ determines the relevant SU(2) as before -  these are of course the SU(2)$_{\text{R}/\text{L}}$ invariant 1-forms. The spinoral Lie derivative\footnote{In general, when taken along a Killing vector $K$, this is defined as
\beq
\mathcal{L}_K\psi= K^m\nabla_m\psi + \frac{1}{4}\nabla_{m}K_{n}\Gamma^{mn}\psi
\eeq
for $\psi$ an arbitrary spinor, $\Gamma_{m}$ a basis curved gamma matrices, $\nabla_{m}$ the covariant derivative.} of the SU(2)$_{\text{L}/\text{R}}$ doublet along the SU(2)$_{\text{L}/\text{R}}$ Killing vectors is 
\beq\label{eq:S2lie}
\mathcal{L}_{K_i}\xi^a = \frac{i\nu}{2} \left(\sigma_i\right)^{a}_{~b} \xi^b,
\eeq 
where $\sigma_i$ are the Pauli matrices, so it is the Lie algebra of SU(2) appearing on the right hand side.
Acting on the SU(2)$_{\text{L}}$ doublet with the SU(2)$_{\text{R}}$ Killing vector on the other hand, or vice-versa, yields zero. One can exploit \eqref{eq:S2lie} to form a spinor transforming in the fundamental of SO(4)$_{\text{L}/\text{R}}\cong $SO(3)$_{1\text{L}/\text{R}}\times$ SO(3)$_{2\text{L}/\text{R}}$ and as a singlet under SO(4)$_{\text{R}/\text{L}}$  depending on the sign of $\nu$.  When one couples such an SO(4) spinor  to an AdS$_3$ spinor as in  \eqref{eq10dspinors}, the result will be a spinor realising the bosonic algebra 
\beq
\mathfrak{s}\mathfrak{l}(2)\oplus \mathfrak{s}\mathfrak{0}(4)
\eeq
as required for large $\mathcal{N}=(4,0)$ supersymmetry. Thus let us now construct such an SO(4) spinor.

In \cite{DeLuca:2018buk} it was established how to form an SO(3) triplet from products of two SU(2) doublets - when the doublets are both formed from $S^3$ Killing spinors, there is only one such triplet (for each sign of $\nu$), namely
\beq
\eta_i= (\sigma_2\sigma_i)_{ab} \xi_1^a\otimes \xi_2^b,
\eeq
it turns out that this is also Majorana.
We define diagonal and anti-diagonal SO(3) Killing vectors as
\beq
K^{+}_i=K^1_i+K^2_i,~~~~K^{-}_i=K^1_i-K^2_i,
\eeq
then it is a simple exercise in Pauli matrix manipulations to establish that
\beq
\mathcal{L}_{K^{+}_i}\eta_j = \nu \epsilon_{ij}^{~~k}\eta_k,~~~~\mathcal{L}_{K^{-}_i}\eta_j =0,
\eeq
so that $K^{+}_i$ realises the Lie algebra of SO(3). We can parameterise a basis for the Lie algebra of SO(4) in block form as 
\beq\label{eq:leialgebra}
(T^+_i)_{ij}=\left(\begin{array}{c|c}\epsilon_{ijk}& \underline{0}\\\hline \underline{0}^T &0\end{array}\right) ,~~~~(T^-_i)=\left(\begin{array}{c|c}0_{3\times 3}& \underline{c_i}\\\hline -\underline{c_i}^T &0\end{array}\right),
\eeq
where $c_1=(1,0,0)^T,~c_2=(0,1,0)^T,~c_3=(0,0,1)^T$.  It is then clear that 3 components of the SO(4) spinor one wishes to construct are simply the SO(3) triplet as these give rise to to the top left blocks of \eqref{eq:leialgebra} under $K^+_i$ and $K^-_i$. The 4th component should be a singlet under the action of $K^{+}_i$ - such a spinor was also provided in \cite{DeLuca:2018buk}, and for $S^3$ is once more unique and Majorana
\beq
\eta_4 = -i(\sigma_2)_{ab} \xi_1^a\otimes \xi_2^b.
\eeq
So there is exactly one SO(4) spinor (for each sign of $\nu$) we can define on $S^3\times S^3$, namely 
\beq
\eta^I=  (\mathcal{M}_I)_{ab} \xi_1^a\otimes \xi_2^b,~~~~\mathcal{M}_I= (\sigma_2 \sigma_1,\sigma_2 \sigma_2,\sigma_2 \sigma_3,-i\sigma_2)_I,
\eeq
It is not hard to confirm that
\beq\label{eq:SO4action}
\mathcal{L}_{K^{\pm}_i}\eta^I= \nu (T^{\pm}_i)^I_{~J}\eta^J, ~~~\eta^{I\dag}\eta^{J}= \delta^{IJ}, 
\eeq
where in the latter we fix an arbitrary normalisation.  
So the spinorial Lie derivative of the SO(4) spinor along the SO(4) Killing vectors realise the associated Lie algebra. We can now write the explicit form of the SO(4) spinors on M$_7$ - given that they must be Majorana, and satisfy $|\chi^I_{1,2}||^2= e^{A}$ component by component  \cite{Dibitetto:2018ftj}, the most general form these can take may be parametrised as
\beq\label{eq:SO4spinors}
\chi^I_1=e^{\frac{A}{2}}\left(\begin{array}{c}\sin(\beta_1+\beta_2)\\i\cos(\beta_1+\beta_2)\end{array}\right)\otimes \eta^I,~~~~\chi^I_2=e^{\frac{A}{2}}\left(\begin{array}{c}\sin(\beta_1-\beta_2)\\i\cos(\beta_1-\beta_2)\end{array}\right)\otimes \eta^I,
\eeq
where $\beta_1,\beta_2$ are functions of $r$ only. Now since each component of $\chi^I_{1,2}$ can be mapped into every other through the action of the R-symmetry \eqref{eq:SO4action}, which we assume the physical fields also respect, we need only explicitly solve the supersymmetry conditions of an $\mathcal{N}=1$ sub-sector to know that $\mathcal{N}=(4,0)$ is preserved. There are several ways to see this but perhaps the most easy argument to follow comes from considering the conditions on \eqref{eq10dspinors} that  follow from setting the gravitino and dilatino variations to zero directly - we defer the proof of this claim to appendix \ref{sec:proof}.

Having established that solving for an $\mathcal{N}=1$ sub-sector of our SO(4) spinor is sufficient to know that $\mathcal{N}=(4,0)$ is realised, we shall now take this sub-sector to be,
\beq\label{eq:neq1sub}
\chi_{1}=\chi^4_{1},~~~~\chi_{2}=\chi^4_{2},
\eeq 
but we stress that as long as the SO(4) R-symmetry is preserved by a solution, this choice is totally arbitrary\footnote{This would no longer be the case if we were to break the R symmetry to some subgroup by for instance fibreing one $S^3$ over the other as in \cite{DeLuca:2018buk} or with the fluxes - then different choices of $\chi_{1,2}$ would lead to different amounts of supersymmetry preserved.}.

Finally, before moving on we should address the issue of a potential enhancement of supersymmetry. If one finds a solution with metric, dilaton and fluxes that do not depend on the signs of $\mu,\nu$ then there exists a second independent 10 dimensional spinor of the form \eqref{eq10dspinors} charged under the second copies of SO(4) and SL(2) at our disposal. This enhances supersymmetry to large $\mathcal{N}=(4,4)$ - as we shall see, this will indeed happen in some instances.

\section{Supersymmetric bi-spinors conditions}\label{sec:two}
Historically one established whether a supergravity solution was supersymmetric by solving spinoral conditions that follow from setting the gravitino and dilatino variations to zero. A more modern approach is the bi-spinor formalism where one instead attempts to solve a set of (generalised) geometric constraints that are necessary and sufficient for supersymmetry. Geometric conditions for AdS$_3$ solutions in type II, i.e. those of the form \eqref{eq:ads3sol}, to preserve $\mathcal{N}=1$ supersymmetry were recently presented in \cite{Dibitetto:2018ftj}. As was done in that reference, we shall impose an additional assumption - that the internal spinors have equal norm which is in 1-to-1 correspondence with $h_0=0$. In IIB this assumption can be made without loss of generality, as all solutions with $h_0\neq 0$ can be mapped to those with $h_0=0$ with SL(2,$\mathbb{R}$) transformations. In IIA it is required for $F_0\neq 0$, but in principle some solutions with non equal norm could exist that we will not see here. 

The fundamental object in the construction of \cite{Dibitetto:2018ftj} is the 7 dimensional bi-spinor $\chi_1\otimes\chi^{\dag}_2$ that is defined in terms of two 7 dimensional Majorana spinors $\chi_{1,2}$ defined on the internal  space M$_7$ as 
\beq\label{eq:bispinor}
\chi_1\otimes\chi^{\dag}_2 = \frac{1}{8}\sum_{n=0}^7\frac{1}{n!} \chi_2^{\dag}\gamma_{a_n...a_1} \chi_1 e^{a_1}\wedge .... \wedge e^{a_n}, ~~~~~ |\chi_1|^2=|\chi_2|^2 = e^{A},~~~~
\eeq
with $\gamma_{a}$ a basis of the flat space gamma matrices in 7 dimensions and $e^a$ is a vielbein on M$_7$ (ie \eqref{eq:vielbein}), as such the bi-spinor is a poly-form. It is a general feature (in odd dimensions) that the bi-spinor can be decomposed as
\beq
\chi_1\otimes\chi^{\dag}_2=\Psi_++ i \Psi_-,
\eeq
for $\Psi_{\pm}$ two real polyforms containing only even/odd forms. An AdS$_3$ solution in type II supergravity of the form \eqref{eq:ads3sol} is guaranteed\footnote{Strictly speaking the following statement is only true away from localised sources. When these are present supersymmetry also requires some additional constraints that we shall discuss when this issue arises.} to satisfy $\mathcal{N}=1$ supersymmetry if it obeys
\begin{align}\label{eq:susycond7d}
&(d-H\wedge)(e^{A-\Phi}\Psi_{\mp})=0,\nn\\[2mm]
&(d-H\wedge)(e^{2A-\Phi}\Psi_{\pm})-2 \mu e^{A-\Phi}\Psi_{\mp}=\pm \frac{e^{3A}}{8}\star_7\lambda(f),\nn\\[2mm]
&e^{-\Phi}(f,\Psi_{\pm})-\frac{\mu}{2} \text{Vol}_7=0,
\end{align}
with the upper/lower signs taken in IIA/B and where $(.~,~.)$ is the Mukai pairing in 7 dimensions defined as
\beq
(X,Y)=  \bigg(\lambda(X)\wedge Y\bigg)_7,
\eeq  
with the operator $\lambda$ defined in \eqref{eq:lambdadef}. The $\mu$ that appears is a constant defining the AdS radius as in \eqref{eq:ads3spinor}, it can in fact be set to $\mu=\pm 1$ by rescaling $e^A$ without loss of generality, but it will become useful to have kept track of it later\footnote{The same applies for $\nu$ and the two 3-sphere warp factor $e^{2C_{1,2}}$ appearing in \eqref{eq:S3KSE} and elsewhere in the previous section.}. 

In the previous section an $\mathcal{N}=1$ spinor was constructed, \eqref{eq:neq1sub}, on M$_7= \mathbb{R}\times S^3\times S^3$ that transforms under an SO(4) R-symmetry. When the physical fields respect this SO(4) the amount of supersymmetry preserved by any solution one can construct from this spinor will be at least $\mathcal{N}=(4,0)$,  with the remaining independent spinors generated through the action of the SO(4) R-symmetry. Specifically one has
\beq
\chi_1=-ie^{\frac{A}{2}}(\sigma_2)_{ab} \left(\begin{array}{c}\sin(\beta_1+\beta_2)\\i\cos(\beta_1+\beta_2)\end{array}\right)\otimes \xi_1^a\otimes \xi_2^b,\quad \chi_2=-ie^{\frac{A}{2}}(\sigma_2)_{ab} \left(\begin{array}{c}\sin(\beta_1-\beta_2)\\i\cos(\beta_1-\beta_2)\end{array}\right)\otimes \xi_1^a\otimes \xi_2^b\nn
\eeq
as this is a tensor product of spinors in each factor of the foliated internal space \eqref{eq:M7} it should be clear from its definition \eqref{eq:bispinor} that the 7 dimensional bi-spinor can be expressed in terms of wedge products of bi-spinors on the interval and two 3-spheres - to this end it is useful to know the bi-spinors on $S^3_{1,2}$ as these are the only non trivial building blocks one requires. One can show, \cite{Macpherson:2017mvu}, that the matrix spinor\footnote{Component by component this is defined as in \eqref{eq:bispinor}, but with $n=0,..3$ and weighted by $\frac{1}{2}$ rather than $\frac{1}{8}$ with the veilbein just the $e^{i_{1,2}}$ parts of \eqref{eq:vielbein}.} following from the two SU(2) spinor doublets of the form \eqref{eq:spinordoublet} are
\beq\label{eq:matrixbispinorS3}
\xi_{1,2}^a \otimes \xi^{b\dag}_{1,2}= \frac{1}{2}\bigg(\big(1- i e^{3C_{1,2}}\text{Vol}(S^3_{1,2})\big)\delta^{ab}+\big(\frac{1}{2}e^{C_{1,2}} K^{1,2}_i-\frac{i}{8}e^{2C_{1,2}}\epsilon_{ijk}K^{1,2}_j\wedge K^{1,2}_k\big)(\sigma^i)^{ab}\bigg)
\eeq
where $e^{2C_{1,2}}$ are the warp factor appearing in \eqref{eq:M7}, and $K^{1,2}_i$ are the SU(2) forms on the two 3-spheres that each obey \eqref{eq:doneform}.  Given \eqref{eq:matrixbispinorS3} it is now a relatively simply exercise to construct $\Psi_{\pm}$. These can be most succinctly written in terms of an SU(3)-structure as
\beq\label{eq:7dbilinears}
\Psi_{+}=\frac{e^{A}}{8}\text{Re}\bigg[e^{i\beta_2} e^{-i J}- e^{k} dr\wedge \Omega\bigg],\quad \Psi_{-}=\frac{e^{A}}{8}\text{Im}\bigg[- e^{i\beta_2}e^{k} dr\wedge  e^{-i J}+\Omega\bigg]
\eeq
where the specific SU(3)-forms are
\begin{align}
J&= \frac{1}{4}e^{C_1+C_2}\bigg(K^1_1\wedge K^2_1+K^1_2\wedge K^2_2+K^1_3\wedge K^2_3\bigg),\nn\\[2mm]
\Omega&=\frac{1}{8}e^{i\beta_1}\big(e^{C_1}K^1_1+i e^{C_2}K^2_1\big)\wedge \big(e^{C_1}K^1_2+i e^{C_2}K^2_2\big)\wedge \big(e^{C_1}K^1_3+i e^{C_2}K^2_3\big).
\end{align}
At this point, in principle, once could blindly plug \eqref{eq:7dbilinears} into \eqref{eq:susycond7d} and find every solution that is consistent with the metric and spinor - but one needs to take a little more care if one wants to ensure that SO(4)$\times$ SO(4) symmetry and  $\mathcal{N}=(4,0)$  supersymmetry is preserved. As long as the dilaton and warp factors of the metric only depend on the interval the only remaining issue is the fluxes. Specifically \eqref{eq:susycond7d} only assumes $\mathcal{N}=1$ supersymmetry is unbroken so the second condition will generically define flux components that break some (super)symmetry. To mitigate this issue we demand that all fluxes must decompose in a basis of the invariant forms of SO(4)$\times$SO(4), namely
\beq
dr,\quad \text{Vol}(S^3_1),\quad \text{Vol}(S^3_2),
\eeq
and their wedge products, with functional support on the interval only - this greatly increases the number of independent conditions in \eqref{eq:susycond7d} that give rise to purely geometric constraints and allows for the exact local form of all solutions consistent with an SO(4)$\times$ SO(4) isometry to be found in the following sections. We study type IIA in section \ref{sec:IIA} and type IIB in in section \ref{sec:IIB}, in both instances we fix the NS 3-form as
\beq
H = c_1 \text{Vol}(S^3_1) +c_2 \text{Vol}(S^3_2),
\eeq
for constants $c_i$ without loss of generality.

\section{All local solutions in type IIA}\label{sec:IIA}
In this section we find all $\mathcal{N}=(4,0)$ solutions with an SO(4)$\times$ SO(4) isometry in type IIA supergravity. There are two independent forms of local solution we study in sections \ref{eq:case1IIA} and \ref{eq:case2IIA} that (generically) preserve $\mathcal{N}=(4,0)$ and $\mathcal{N}=(4,4)$ respectively. We show how the former can be used to construct new compact global solutions, and provide a hint as to how one might do the same with the latter.

Upon plugging the bi linears of \eqref{eq:7dbilinears} into \eqref{eq:susycond7d} one quickly realises two zero form constraints
\beq
(\cos\beta_1 e^{C_1}-\sin\beta_1 e^{C_2})=(\mu \cos\beta_1 e^{C_1}-\nu e^{A}\sin\beta_2)=0\label{eq:SUSYIIA0}
\eeq
these are very useful as they cannot be solved when any of $\cos\beta_1$, $\sin\beta_1$ or $\sin\beta_2$ are set to zero, as this would require us to do the same to one of the warp factors. We can then take \eqref{eq:SUSYIIA0} as  general definitions for $e^{C_i}$ in IIA and eliminate these factors from the rest of the supersymmetry constraints, after some work we find the additional conditions
\begin{align}
&\beta_1'=F_2=F_0\cos\beta_2=0,\label{eq:SUSYIIA1}\\[2mm]
&c_1 \cos^4\beta_1+c_2\sin^4\beta_1=\mu^2 c_2\sin^3\beta_1+\nu^3\cos\beta_1 e^{2A}\sin(2\beta_2)=0,\label{eq:SUSYIIA2}\\[2mm]
&(e^{5A-\Phi}\sin^3\beta_2)'-2\mu e^{4A+k-\Phi}\cos\beta_2 \sin^2\beta_2=0,\label{eq:SUSYIIA3}\\[2mm]
&(e^{3A}\sin^3\beta_2)'-\frac{3\mu}{2}e^{2A+k}\cos\beta_2+ \frac{3}{4}e^{3A+k-\Phi}\sin^2\beta_2F_0=0,\label{eq:SUSYIIA4}
\end{align}
where the last of these comes from imposing that $F_0$ is constant - the rest of the Bianchi identities then follow rather trivially.
Clearly there are two cases, $F_0=0$ and $\cos\beta_2=0$.

\subsection{Case I: compact solutions from D8/O8s back reacted on AdS$_3\times S^3\times S^3\times S^1$}\label{eq:case1IIA}
For Case I we set
\beq
\cos\beta_2=0,\quad \sin\beta_2= s,\quad s=\pm 1,
\eeq
then \eqref{eq:SUSYIIA2} implies also
\beq
c_1=c_2=0,
\eeq
so there is no NS flux turned on. We can solve \eqref{eq:SUSYIIA3} by fixing
\beq
e^{5A-\Phi}=q  L^5,
\eeq
where $L$ and $q$ are constants and we use diffeomorphism invariance to fix
\beq
e^{A+k}=q L.
\eeq
We can use \eqref{eq:SUSYIIA0} to define $e^{C_1},e^{C_2}$ which leaves \eqref{eq:SUSYIIA4} to solve. This becomes simply $(L^4e^{-4A})'=\nu F_0$, which is solved by
\beq
L^4e^{-4A}=  H_{8},\quad H_8= F_0 \nu r+ c
\eeq
for $c$ another constant - i.e. the warp factor of a D8 brane or O8 hole depending on the sign of $F_0$ and $\nu$. We then fix
\beq
s=\mu=\nu=\pm 1
\eeq
and find the following general form for local solutions in massive IIA
\begin{align}\label{eq:sol1}
ds^2&=\frac{1}{\sqrt{H_{8}}}\bigg(L^2 ds^2(\text{AdS}_3)+\frac{L^2}{\cos^2\beta_1}ds^2(S^3_1)+\frac{L^2}{\sin^2\beta_1}ds^2(S^3_2)\bigg)+ \sqrt{H_{8}} q^2 dr^2,\nn\\[2mm]
F_4&= 2 q^2 H_{8}\bigg(L^2\text{Vol}(\text{AdS}_3)+ \frac{L^2}{\cos^2\beta_1}\text{Vol}(S^3_1)+\frac{L^2}{\sin^2\beta_1}\text{Vol}(S^3_2)\bigg)\wedge dr,\nn\\[2mm]
e^{-\Phi}&=q H^{\frac{5}{4}}_8.
\end{align}
Clearly when $F_0=0$ we recover the standard solution on $AdS_3\times S^3\times S^3\times S^1$ which preserves $\mathcal{N}=(4,4)$ supersymmetry - in this limit none of the physical fields depend on $\mu=\nu=\pm 1$ confirming the enhanced supersymmetry. The generic local solution is D8 branes or O8 planes or both\footnote{The near horizon geometry of a D8 brane is indistinguishable from the  geometry near an O8 plane.} back reacted on this. As the warp factor now depends on $\nu$  supersymmetry is broken to $\mathcal{N}=(4,0)$ in the presence of the back reacted D8/O8 system - which is by no means a surprise.

When $ F_0\neq 0 $ the internal space of \eqref{eq:sol1} is non compact, if we assume $F_0>0$ and $\nu=1$, then the warp factor does bound the interval from below at $r=-\frac{c}{F_0}$ where the behaviour is consistent with a D8/O8 system wrapped on AdS$_3\times S^3 \times S^3$, however the interval is not bounded from above and  $r\to\infty$ is at infinite proper distance. Before giving up though, one should remember that this is only a local solution - which is to say that all coordinate patches of a global solution can be expressed in the form \eqref{eq:sol1}. One can try to make a compact solution by gluing a second mirrored copy of \eqref{eq:sol1} onto the first in the spirit of 	\cite{Apruzzi:2013yva}. At the point where the local patches connect there should be a D8 brane defect where $F_0$ jumps, but the metric and dilaton are continuous. The simplest way to arrange for this is to glue the patches together at $r=0$ and have $F_0$ flip from positive to negative as one crosses $r=0$ from below, i.e. one takes the warp factor to be
\beq\label{eq:Hmod}
H_{8}= c+ |F_0|r, ~~~r<0,~~~~~~~~~~~~~~~~~~~H_{8}= c- |F_0|r, ~~~r>0,
\eeq
so that the metric and dilaton are continuous without the need to  further tune constants, and only $F_0$ jumps. This does indeed bound $r$ to the interval $\mathcal{I}$ between two D8/O8 systems at $r= \pm \frac{c}{|F_0|}$ and one is now able to quantise the fluxes without issue. In units where $g_s=\alpha'=1$ one requires that the following charges are integer valued
\beq
n_0=2\pi F_0 ,~~~ N_{2} =-\frac{1}{(2\pi)^5}\int_{S^3_1\times S^3_2}\star F_4,~~~ N^i_{4} =\frac{1}{(2\pi)^3}\int_{S^3_i}\int_{r\in\mathcal{I}}F_4.
\eeq
This is not hard to achieve by tuning 
\beq
\frac{2 \pi F_0}{c_1^2 q^2 L^2}= \frac{1}{N^1_4}+ \frac{1}{N^2_4},~~~~\frac{L^5 q}{4\pi\sin^3\beta_1\cos^3\beta_1}= N_1,~~~~\tan\beta_1= \frac{N^2_4}{N^1_4},
\eeq
and the curvature of the solution is under parametric control. A standard computation\footnote{The holographic central charge for a warped AdS$_3$ solution in 10 dimensions at leading order is given for instance in \cite{Couzens:2017way}, converting this reference to string frame and setting $\alpha'=g_s=\mu=1$ gives 
\beq
c= \frac{3}{2^4 \pi^6}\int_{M_7}e^{A-2\Phi}\text{Vol}(\text{M}_7).
\eeq
Using this formula with $F_0=0$ (ie the AdS$_3\times S^3\times S^3\times S^1$ limit) one finds  $c= 6N_1\frac{N^1_4 N^2_4}{N^1_4+ N^2_4}$ which implies  $k_+= N_1 N^1_4$, $k_-= N_1 N^2_4$ and $\alpha=\tan^2\beta_1= \frac{N^4_2}{N^4_1}$.} then leads to the finite central charge
\beq
c = 6 c_1  N_2 \frac{N_4^1N_4^2}{N_4^1+ N_4^2}+ {\cal O} (1)
\eeq
where ${\cal O}(1)$ parametrises sub leading terms that supergravity is insensitive to (at least with the computation performed here). Note that this is independent of $F_0$ and when $c_1=1$ is actually the same central charge as that of AdS$_3\times S^3\times S^3\times S^1$. More generally it is consistent with \eqref{eq:chargealpha}, i.e. what one expects from a CFT with large superconformal algebra provided $c_1$ is integer - indeed one can identify
\beq
k_+= c_1 N_2 N^1_4,~~~~k_-= c_1 N_2 N^2_4,~~~~ \alpha = \tan\beta_1^2 = \frac{N^4_2}{N^4_1},
\eeq 
with $\alpha$ the continuous parameter of $\mathfrak{d}(2,1,\alpha)$. 

This all sounds very positive, however to be sure this solution really exists, and preserves supersymmetry, we need to check the Bianchi identities at the D8 brane defect are satisfied and that the D8 brane is calibrated  \cite{Martucci:2005ht,Lust:2010by,Legramandi:2018qkr}. The Bianchi identities in the presence of the defect require
\beq\label{eq:bianchi}
d f= \frac{N_8}{2\pi} \delta(r)e^{2 \pi f_{g}}\wedge dr 
\eeq 
where $f_{g}$ is a gauge field on the world volume of the D8 brane. We find ourselves in a far simpler scenario than \cite{Apruzzi:2013yva}, because there is no NS flux and the only RR flux that shifts across the defect is $F_0$ as
\beq
\Delta F_0= 2|F_0|.
\eeq
Comparing this with the integrated form of \eqref{eq:bianchi}, we find that the Bianchi identity requires simply
\beq
N_8= 4\pi|F_0|= 2|n_0|,\quad f_{g}=0.
\eeq
It is also not hard to confirm that the brane is supersymmetric - this is so whenever the DBI action  of a given brane satisfies a so called calibration condition. Here the DBI action of the D8 should equal the integral of $e^{3A-\Phi}\text{Vol}(\text{AdS}_3)\wedge\Psi_6$ - a quick computation shows this to indeed be the case. Thus we have constructed a bona fide $\mathcal{N}=(4,0)$ solution in massive IIA.

The result of gluing the two local solutions together is essentially a global solution with an orientifold under which a circle parameterised by $r$ becomes a segment. At the two ends there are two O8-planes with different charges and tensions. One can interpret this as two O8$_-$s with  $k$  and $16-k$ D8s, or when $k=8$ as an O8$_-$ and an O8$_+$, the later being similar to what appears in the recently constructed classical dS solutions in \cite{Cordova:2018dbb}. 

It would be interesting to find the local solution of \eqref{eq:sol1} as a near horizon limit of some intersecting brane set-up. This should be in some sense a trivial extension of the realisation of AdS$_3\times S^3\times S^3\times S^1$ in terms of D2 and D4 branes -  however to the author's knowledge this first step is currently absent from the literature (see \cite{deBoer:1999gea} for every near horizon realisation of AdS$_3\times S^3\times S^3\times S^1$ except D2-D4), and finding it is beyond the scope here.

Finally let us stress that there is not any particular need to place the D8 brane defect at $r=0$ and so there is  no restriction to gluing together an arbitrary number of local solutions of the form \eqref{eq:sol1}, with a D8 brane defect at each intersection in the vein of \cite{Cremonesi:2015bld}. All one needs to ensure is that $h_8$ is continuous across each defect by tuning $c_1,F_0$ and the intersection points in each local patch, and that the interval has an upper and lower bound. One can construct infinitely many globally distinct compact solutions in this fashion, so it would be interesting to study this possibility in more detail. The dual CFT interpretation of this infinite class will presumably be adding various amounts of conformal matter to the CFT dual of AdS$_3\times S^3\times S^3\times S^1$ in such a way that $\mathcal{N}=(4,4)$ is broken to $\mathcal{N}=(4,0)$ - but that remains to be seen. 

\subsection{Case II: a new local $\mathcal{N}=(4,4)$ solution with O2 Plane}\label{eq:case2IIA}
For Case II we set
\beq
F_0=0.
\eeq
To avoid falling into a sub-case of the previous section one must demand $\cos\beta_2\neq 0$ which requires the same of $c_1,c_2$ without loss of generality. We can solve the first condition of  \eqref{eq:SUSYIIA2} with
\beq
c_1=c \sin^4 \beta_1,\quad c_2= -c \cos^4 \beta_1
\eeq
and take the second condition as the definition of $e^{A}$. Using this, and by taking a linear combination of  \eqref{eq:SUSYIIA3} and \eqref{eq:SUSYIIA4} such that $e^{k}$ is eliminated one finds $(\tan\beta_2e^{-2\Phi})'=0$ which is solved by
\beq
\tan\beta_2 e^{-2\Phi}= q^2
\eeq
where $q$ is a constant. At this point it is useful to use diffeomorphism invariance to fix $e^k$ in terms of another arbitrary function $f(r)$ such that
\beq
32 \nu^3e^{A+k}= c\mu \sin^3\beta_1 f'
\eeq
the remaining conditions \eqref{eq:SUSYIIA3}-\eqref{eq:SUSYIIA4} both then reduce to $f'= (\sec\beta_2)'$ which one can solve without loss of generality with
\beq
\cos\beta_2= \frac{1}{\sqrt{f}}.
\eeq
As the left hand side of this expression is bounded between 0 and 1, we must have that $1<f<\infty$, a sensible choice is then
\beq
f= \frac{1}{\cos^2 r},
\eeq
so that we simply have $\beta_2=r$. This leads to a completely determined local solution of the form
\begin{align}\label{eq:sol2}
ds^2&=L^2\bigg[ \frac{1}{\cos r \sin r}ds^2(\text{AdS}_3)+ \frac{\sin^3 r}{\cos^5 r}dr^2+\frac{\sin r}{\cos r}\bigg(\frac{1}{\cos^2\beta} ds^2(S^3_1)+ \frac{1}{\sin^2\beta}ds^2(S^3_2)\bigg)\bigg],\nn\\[2mm]
H&=2L^2\bigg(\frac{\tan\beta}{\cos^2\beta}\text{Vol}(S^3_1)-\frac{\cot\beta}{\sin^2\beta}\text{Vol}(S^3_2)\bigg),\quad q e^{\Phi}= \sqrt{\tan r},\\[2mm]
F_4&= 2 L^3\bigg[4 \frac{\sin 4r}{\sin^2 2r} \text{Vol}(\text{AdS}_3)+q\frac{\tan r}{\cos^2r}\bigg(\frac{1}{\cos^2\beta}\text{Vol}(S^3_1)+\frac{1}{\sin^2\beta}\text{Vol}(S^3_2)\bigg)\bigg]\wedge dr,\nn
\end{align}
where we have introduced
\beq
L^2= \frac{c}{2} \cos^3\beta \sin^3\beta,
\eeq
and fixed
\beq
\sin\beta_1= \nu \sin\beta, \quad \mu=\nu =\pm 1,
\eeq
to simplify expressions. Notice that none of the physical fields depend on $\nu=\pm 1$ - so this solution experiences an enhancement of supersymmetry to large $\mathcal{N}=(4,4)$.

The internal radial coordinate is bounded as $0<r<\frac{\pi}{2}$, with the lower bound a singularity of the metric. The behaviour close to $r=0$ is intriguing, indeed after redefining $r=\sqrt{y}$ the behaviour is that of O2 planes at the base of a cone over $S^3\times S^3$, which is rather novel. More disappointing is the behaviour close to $r=\frac{\pi}{2}$ where the metric is actually regular but the dilaton is infinite, which does not appear to be physical behaviour. Worst still perhaps, is that $r=\frac{\pi}{2}$ is at infinite proper distance, so the internal space is non-compact. One way to see this is with the central charge which goes like
\beq
c \sim \lim_{r\to \frac{\pi}{2}} \tan^4 r
\eeq
which  is clearly divergent. Thus any putative CFT dual will have a continuous operator spectrum, a sign that it is sick.

One might be able to cure this issue as before by gluing two copies of \eqref{eq:sol2} together. As $F_0=0$, one can no longer achieve this with D8 branes. However, since this sort of gluing does work with D8 branes, T-duality and S-duality informs us that at the very least, it should be possible to glue solutions together with other types of branes when they are smeared over all but one of their co-dimensions - the options here are D2 and NS5 branes. As this may be a way of constructing  new holographic duals to well defined CFTs with large $\mathcal{N}=(4,4)$ supersymmetry it would certainly be interesting to peruse this possibility in future.

Finally, since this solution has no Romans mass turned on, it can be lifted to M-theory. As such this solution fits within the classification of \cite{DHoker:2008lup,Estes:2012vm,Bachas:2013vza}. At first sight this sheds doubt on the possibility of constructing compact solutions by gluing copies of \eqref{eq:sol2} together with defects, as \cite{Bachas:2013vza} claims that the only AdS$_3$ solutions in M theory with compact internal space are locally AdS$_3\times S^3\times S^3 \times T^2$. This follows from ruling out the possibility of compact solutions with localised sources on the boundary of the Riemann surface orthogonal to AdS$_3\times S^3\times S^3$, however senarios with addtional defects on the interior do not appear to have be considered in \cite{Bachas:2013vza}. Thus if compact solutions can be realised from \eqref{eq:sol2} it is then possible that a broader class of  $\mathcal{N}=(4,4)$ solutions with defects may also exist in M-theory - but at this point this is just speculation.

\section{All  local solution in type IIB}\label{sec:IIB}
In this section we find the local form of all solutions preserving at least an SO(4) R-symmetry on  $S^3\times S^3$. In section \eqref{eq:case1IIB} we find a new compact solution containing D5s and O5s which actually preserves small $\mathcal{N}=(4,0)$ supersymmetry, while in section \eqref{eq:case2IIB} we find a solution that back reacts D5s on AdS$_3\times S^3 \times S^3 \times S^1$, but is non compact.

Once again we begin by plugging the bi linears of \eqref{eq:7dbilinears} into \eqref{eq:susycond7d}. It is immediate to establish the following zero form constraints
\beq
c_1\cos\beta_2=c_2\cos\beta_2=\mu\sin\beta_2=0
\eeq
which means that for AdS$_3$ solutions we must set
\beq
c_1=c_2= \sin\beta_2=0
\eeq
and so all flux components but $f_3$ and $f_7$ are set to zero. We will thus parametrise the 3-form in terms of two constants $c_3,c_4$ as
\beq
f_3= c_3\text{Vol}(S^3_1)+c_4 \text{Vol}(S^3_2).
\eeq
Given this, and after some massaging of expressions, it is possible to extract the following algebraic constraints 
\begin{align}
&\cos\beta_2= s,\\[2mm]
&\nu e^{A}(\cos\beta_1 e^{C_1}-\sin\beta_1 e^{C_2})- s\mu e^{C_1+C_2}=0,\label{eq:case1IIBa}\\[2mm]
& e^{A}(c_3\sin\beta_1 e^{3C_2} -c_4 \cos\beta_1 e^{3C_1})+2 \mu e^{3C_1+3C_2-\Phi}=0\label{eq:case1IIBb},
\end{align}
where $s=\pm 1$. Using these we can simplify the  differential constraints to
\begin{align}
&(\beta_1)'=\cos\beta_1\sin\beta_1(e^{C_1-C_2})'=0,\label{eq:case11IIBa}\\[2mm]
&(e^{2A+2C_1+2 C_2-\Phi})'=2 s\nu e^{2A+C_1+C_2+k-\Phi}(\cos\beta_1 e^{C_2}+\sin\beta_1 e^{C_1}),\label{eq:case11IIBb}\\[2mm]
&(e^{3A+2C_1+C_2-\Phi}\cos\beta_1)'=2 e^{2A+C_1+C_2+k-\Phi}(s \nu e^{A}+\mu \sin\beta_1e^{C_1}),\label{eq:case11IIBc}\\[2mm]
&(e^{3A+C_1+2C_2-\Phi}\sin\beta_1)'=2 e^{2A+C_1+C_2+k-\Phi}(s \nu e^{A}-\mu \cos\beta_1e^{C_2}),\label{eq:case11IIBd}\\[2mm]
&(e^{3A+3C_2-\Phi}\cos\beta_1)'- 2 \mu e^{2A+3C_2+k-\Phi}\sin\beta_1=c_3 e^{3A-3C_1+3C_2+k},\label{eq:case11IIBe}\\[2mm]
&(e^{3A+3C_1-\Phi}\sin\beta_1)'+ 2 \mu e^{2A+3C_1+k-\Phi}\cos\beta_1=c_4 e^{3A+3C_1-3C_2+k},\label{eq:case11IIBf}
\end{align}
which are not all independent, but this form makes finding a solution easier.

It appears that there are 3 cases, $\cos\beta_1=0$, $\sin\beta_1=0$ and $(e^{C_1-C_2})'=0$, however there is no physical difference between the first two of these as one is mapped to the other by relabelling the spheres - thus there are two physically distinct cases.

\subsection{Case I: a new compact solution with  D5s and O5s back reacted on AdS$_3\times S^3\times \mathbb{R}^4$}\label{eq:case1IIB}
For Case I we set
\beq
\cos\beta_1= t=\pm 1
\eeq
and take \eqref{eq:case1IIBa}-\eqref{eq:case1IIBb} to define the AdS warp factor and dilaton. Substituting this back into \eqref{eq:case11IIBa}-\eqref{eq:case11IIBf} one  is left with just  two independent conditions
\beq\label{eq:caseIIIBeq}
s\nu e^{C_2+k}- t (e^{C_1+C_2})'=0,~~~~\nu c_3 e^{k}+ c_4 s t e^{3C_1}(e^{-2C_2})'=0.
\eeq
We use diffeomorphism invariance to fix
\beq
\nu e^{k+C_2}=s t,
\eeq
which trivialises the first ODE of \eqref{eq:caseIIIBeq}, and so allows us to integrate both as
\beq
e^{C_1+C_2}= L r,~~~~~ L^2e^{-4C_2}=h_5= a+ \frac{c_3}{c_4 r^2},
\eeq
for $a$ an arbitrary constant and where we now fix $s=t=\mu=\nu=\pm 1$. The resulting solution then takes the form
\begin{align}\label{eq:metfluxd5}
ds^2&= L^2\bigg[\frac{1}{\sqrt{h_5}}\bigg(ds^2(\text{AdS}_3)+ ds^2(S^3_2)\bigg)+\sqrt{h_5}\bigg(dr^2+r^2 ds^2(S^3_1)\bigg)\bigg],~~~~e^{-\Phi}=\frac{c_4}{2L}\sqrt{h_5},\nn\\[2mm]
F_3&= c_4\bigg( \text{Vol}(\text{AdS}_3)+\text{Vol}(S^3_2)\bigg)+ \nu c_3 \text{Vol}(S^3_1).
\end{align}
When $a=1$ and $\frac{c_3}{c_4}>0$, $h_5$ is the warp factor of a D5 brane so the solution is  D5 branes back reacted on AdS$_3\times S^3\times \mathbb{R}^4$, which is non compact. When $c_3=0$ the solution no longer depends on $\nu$ and so supersymmetry is enhanced to $\mathcal{N}=(4,4)$, consistent with the fact that the solution is locally  AdS$_3\times S^3\times T^4$ in this limit. 

It may be possible to construct a globally compact solution by gluing copies of \eqref{eq:metfluxd5} together with D5 brane defects in a similar fashion to section \ref{eq:case1IIA}, however for this case there is an easier way to achieve this. The previous discussion depended on tuning $a,c_3,c_4$ in a certain fashion, but there is no requirement to do this - indeed if one assumes $ a<1$ then $r$ becomes bounded to the interval $[0,~\sqrt{\frac{c_3}{c_4 |a|}}]$ and the solution is compact. To see this more clearly one can perform the coordinate transformation and redefinition
\beq
r\to \sqrt{\frac{c_3}{|a|c_4}}\cos r,~~~~ L^2\to  |a| L^2,
\eeq
which modifies the metric and dilaton as
\begin{align}\label{eq:compsol}
ds^2&= L^2\bigg[\frac{\cos r}{\sin r}\bigg(ds^2(\text{AdS}_3)+ ds^2(S^3_2)\bigg)+\frac{c_3 \sin r}{c_4 \cos r}\bigg(\sin^2 r dr^2+\cos^2 r ds^2(S^3_1)\bigg)\bigg],\nn\\[2mm]
F_3&= c_4\bigg(  \text{Vol}(\text{AdS}_3)+\text{Vol}(S^3_2)\bigg)+ \nu c_3 \text{Vol}(S^3_1),\quad e^{-\Phi}=\frac{c_4}{2L^2} \tan r,
\end{align}
and leaves the flux unchanged. Clearly $c_3$ can no longer be set to zero so the  $\nu$ dependence of the flux means that only $\mathcal{N}=(4,0)$ supersymmetry is preserved in general. The internal radius is now bounded as $0<r<\frac{\pi}{2}$ and at the end points there are singularities, however these have an obvious physical origin. It should not be hard to see that close to $r=0$ the metric becomes that of O5 planes wrapped on AdS$_3\times S^3_2$, while at $r=\frac{\pi}{2}$ it  behaves as  D5s wrapped on AdS$_3$ and either of the two 3-spheres.  

Flux quantisation requires that the following charges are integer
\beq
N^i_{5} =  \frac{1}{(2\pi)^2}\int_{S^3_i}F_3,\quad N_1 =-\frac{1}{(2\pi)^6}\int_{\mathbb{R}\times S^3_1\times S^3_2}\star F_3.
\eeq
This can be simply achieved by tuning
\beq
c_{i+2}=2 N^i_{5},\quad \frac{c_3^2 L^4}{64 c_4 \pi^2}  = N_1,
\eeq
and the radius about the singularities for which the supergravity approximation does not hold can be made arbitrarily small by making $L$ (and so the D1 charge) large.  We once more compute the holographic central charge at leading order and find that it is finite with the following form 
\beq\label{eq:centralcharge2}
c= 6 N_1 N^2_5+ {\cal} O(1).
\eeq
This behaviour is markedly different from that of AdS$_3\times S^3\times S^3 \times S^1$, indeed the form the central charge takes is consistent with small $\mathcal{N}=(4,0)$ superconformal algebra  with level $k=N_1 N^2_5$. Since this solution is  D5s and O5s back reacted on AdS$_3\times S^3\times \mathbb{R}^4$, which itself preserves (two copies of) small $\mathcal{N}=(4,0)$, this should not be surprising. It is of course well known that small $\mathcal{N}=(4,0)$ comes equipped with only an SU(2) R-symmetry which at first sight may appear at conflict with the SO(4) R-symmetry preserving spinors from which this solution is constructed. In fact this apparent conflict exists for AdS$_3\times S^3\times \mathbb{T}^4$ as well and the resolution is the same for both cases. The SO(4) R symmetry of the geometry is actually realising both the SU(2) R-symmetry of the dual CFT and an SU(2) outer automorphism symmetry\footnote{I thank the reviewer for clarification on this point} and the dual CFT indeed has small superconformal algebra. Such solutions lie within the $\alpha\to 0$ limit of $\mathfrak{d}(2,1,\alpha)$ superconformal symmetry which is isomorphic to $\mathfrak{psu}(2,1,\alpha)\rtimes \mathfrak{su}(2)$, so \eqref{eq:compsol} lies within a degenerate limit of large $\mathcal{N}=(4,0)$.   .

This solution is a well defined AdS dual to an as yet to be determined CFT$_2$ with small  $\mathcal{N}=(4,0)$ superconformal symmetry so is well deserving of further detailed study - but this is beyond the scope here.

\subsection{Case II: D5s back reacted on AdS$_3\times S^3\times S^3\times S^1$}\label{eq:case2IIB}
In case II we assume 
\beq
0<\sin\beta_1<1,\quad s=\mu =\nu=\pm 1.
\eeq
Due to \eqref{eq:case11IIBa} this means that the two 3-sphere warp factors can only differ by a constant, thus we introduce a new function, $H$ and constants $b_1,b_2$ such that
\beq
e^{C_i}= b_i H.
\eeq
We again use \eqref{eq:case1IIBa}-\eqref{eq:case1IIBb} as definitions for $e^{A},e^{\Phi}$ and substitute for these quantities in \eqref{eq:case11IIBa}-\eqref{eq:case11IIBf} - they once more reduce to just two conditions that may be easily solved with
\beq
b_1= b \sqrt{c_3},\quad b_2= b \sqrt{c_4},\quad \nu e^{C+k}= b q ,\quad H= c+ \nu \lambda_1 r,
\eeq
where we introduce
\beq
\lambda_1= \frac{q \cos\beta_1}{\sqrt{c_3}}-\frac{q \sin\beta_1}{\sqrt{c_4}},\quad \lambda_2= \frac{\cos\beta_1}{\sqrt{c_4}}+\frac{\sin\beta_1}{\sqrt{c_3}},
\eeq
to ease notation and $b,c$ are constants. The general local form of solutions can then be written as 
\begin{align}\label{eq:sol4}
ds^2&= L^2 H\bigg(ds^2(\text{AdS}_3)+ c_3 \lambda_2 ds^2(S^3_1)+c_4 \lambda_2 ds^2(S^3_2)\bigg)+ \frac{L^2 q^2 \lambda_2^2}{H}dr^2,\nn\\[2mm]
F_3&=\frac{1}{\lambda_2^2}\text{Vol}(\text{AdS}_3)+c_3\text{Vol}(S^3_1)+ c_4 \text{Vol}(S^3_2),\quad e^{\Phi}=2 L^2 \lambda_2^2 H,~~~~b= L \lambda_2
\end{align}
The warp  factor $H$ depends on $\nu$ so this solution generically experiences no enhancement beyond $\mathcal{N}=(4,0)$. Similar to section \ref{eq:case1IIA} however, when one sets $\lambda_1=0$, $\nu$ drops out of all expressions and supersymmetry is enhanced to $\mathcal{N}=(4,4)$ - this is because the solution becomes AdS$_3\times S^3\times S^3\times S^1$ in this limit.  The generic solution has D5 branes back reacted on this. The attentive reader will note that  $H$ is not the warp factor of a D5 brane, however, if one assumes $\nu=1$, then the interval is bounded from below at 
\beq
r= -\frac{c}{\lambda_1},
\eeq
where the near horizon geometry of a D5 brane wrapped on either $S^3$ is recovered. The interval is not however bounded from above and $r=\infty$ is at infinite proper distance, so the metric is non compact.

One might wonder about the possibility of making the solution compact by gluing two copies of \eqref{eq:sol4} together with D5 branes smeared on $S^3$ - the issues are essentially the same as for Case II in IIA.

\subsection*{Acknowledgements}
I would like to thank Alessandro Tomasiello, Gabriele Lo Monaco and Ida Zadeh for useful discussions.  I am funded by the Italian Ministry of Education, Universities and Research under the Prin project ``Non Perturbative Aspects of Gauge Theories and Strings'' (2015MP2CX4) and INFN. 

\appendix

\section{Conventions for sections \ref{sec:one} and \ref{sec:two}}\label{sec:conventions}
In this section we detail a set of conventions that can be used to perform the calculations in sections \ref{sec:one} and \ref{sec:two} - one should understand however that the result of these sections do not strictly depend on this choice, which is why they do not appear in the main text.
\subsection{Spinors and gamma matrices on AdS$_3\times \text{M}_7$}
We decomposing the 10 dimensional gamma matrices as in \cite{Haack:2009jg}
\beq
\Gamma_{\mu}= \gamma^{\text{AdS}_3}_{\mu}\otimes \sigma_3 \otimes \mathbb{I},~~~\Gamma_a= \mathbb{I}\otimes \sigma_1 \otimes \gamma_a,~~~ B^{10}= \mathbb{I}\otimes \sigma_3\otimes B,~~~ \hat\Gamma =- \mathbb{I}\otimes \sigma_2 \otimes \mathbb{I},
\eeq
where $\gamma^{\text{AdS}_3}_{\mu}$ are a real basis of flat space gamma matrices on AdS$_3$, $\gamma_a$ a 7 dimensional basis such that $B^{-1}\gamma_a B=- \gamma_a^*$ and $B B^*= \mathbb{I}$, $\sigma_{1,2,3}$ are the Pauli matrices and $\hat\Gamma$ is the chirality matrix. A 10 dimensional spinor of $\pm$ chirality on AdS$_3\times \text{M}_7$ can then be decomposed in terms of this basis as
\beq
\epsilon=\zeta\otimes v_{\pm} \otimes \chi,~~~~~~~~~~ v_{\pm} = \left(\begin{array}{c}1 \\ \mp i\end{array}\right),
\eeq 
with $\zeta$ a real Killing spinor on AdS$_3$, $\chi$ a spinor on M$_7$ and $v_{\pm}$ an auxiliary 2 vector which  takes care of 10 dimensional chirality as $-\sigma_2 v_{\pm} = \pm v_{\pm}$ and is required to make a representation of the gamma matrices on AdS$_3\times \text{M}_7$ 32 dimensional. If $\epsilon$ is one of the Killing spinor for type II supergravity it should be Majorana, here this just requires imposing that $\chi$ is Majorana, ie
\beq
\chi^c= B \chi^* =\chi,
\eeq
as $\zeta$ are real and $v_{\pm}=\sigma_3 v_{\pm}^*$. This is what leads to the form of the 10 dimensional Killing spinors taken in \eqref{eq10dspinors}. 
\subsection{Vielbein and gamma matrices in on M$_7$} 
The internal space M$_7$ is  a foliation of $S^3\times S^3$ over an interval as in \eqref{eq:M7}, we can define a veilbein on this space as
\beq\label{eq:vielbein}
e^r = e^k dr,~~~ e^{i_{1,2}} = e^{C_{1,2}} \hat e^{i_{1,2}},~~~ \hat e^{i_{1,2}} = \frac{1}{4}(d\theta_{1,2},~\sin\theta_{1,2} d\phi_{1,2},~ d\psi_{1,2}+ \cos\theta_{1,2} d\phi_{1,2})^{i_{1,2}},   
\eeq
where $\hat e^{i_{1,2}}$ are in the Hopf fibration frame mentioned before \eqref{eq:spinordoublet} in the main text. A suitable basis of 7 dimensional gamma matrices are then
\beq\label{eq:7dgammas}
\gamma_{r} = e^{k}\sigma_1\otimes \mathbb{I}\otimes \mathbb{I},~~~\gamma_{i_1} = e^{C_1}\sigma_2\otimes \gamma^{S^3_1}_{i_1}\otimes \mathbb{I},~~~\gamma_{i_2} = e^{C_2}\sigma_3\otimes \mathbb{I}\otimes \gamma^{S^3_2}_{i_2},~~~ B= \sigma_1\otimes \sigma_2\otimes \sigma_2,
\eeq
where $\gamma^{S^3_{1,2}}_{i_1}$ are the unwarped gamma matrices of $S^{3}_{1,2}$ that are flat with respect to $\hat e^{i_{1,2}}$. Euclidean gamma matrices in 3 dimensions are always the Pauli matrices, up to signs and ordering, so we can in fact take
\beq\label{eq:3dgammas}
\gamma^{S^3_{1}}_{i_{1}}= \sigma_{i_{1}},~~~~\gamma^{S^3_{2}}_{i_{2}}= \sigma_{i_{2}}
\eeq
without loss of generality. 
\subsection{Conventions on $S^3$}
The SO(4) spinors we construct in \eqref{eq:SO4spinors} are a certain product of Killing spinors on $S^3_{1,2}$ that transform as  \eqref{eq:SO4action} along the sum and difference of the SU(2)$_{\text{L}/\text{R}}$ Killing vectors of each 3-sphere. Here we will be more explicit about the form these Killing spinors/vectors in terms of \eqref{eq:vielbein} - \eqref{eq:3dgammas}. To this end, let us here label an objects dependence on SU(2)$_{\text{L}/\text{R}}$ explicitly - we will focus on an arbitrary unwarped 3-sphere of unit radius in the Hopf fibration frame
\beq\label{eq:hopfframe}
\hat e^i= \frac{1}{4}(d\theta,~\sin\theta d\phi,~ d\psi+ \cos\theta d\phi)^{i},
\eeq
which can be related to the main text by simply adding an index where appropriate. The 3-sphere Killing spinor equations \eqref{eq:S3KSE} then becomes
\beq
\nabla_{i}\xi^{\text{L}}= \frac{i}{2}\sigma_{i} \xi^{\text{L}},~~~~\nabla_{i}^{S^3}\xi^{\text{R}}=- \frac{i}{2}\sigma_{i} \xi^{\text{R}},
\eeq
which are solved in general by
\beq
\xi^{\text{L}}= e^{\frac{i}{2}\theta \sigma_1}e^{\frac{i}{2}\phi \sigma_3}\xi^0,~~~~~\xi^{\text{R}}= e^{-\frac{i}{2}\psi \sigma_3}\xi^0,
\eeq
for $\xi^0$ a constant spinor.  We fix  $\sigma_3\xi^0=\xi^0$ without loss of generality. The SU(2)$_{\text{L}/\text{R}}$ Killing vectors are given by
\begin{align}
	K^{\text{R}}_1+i K^{\text{R}}_2 &= e^{i\psi}(i\partial_{\theta} + \csc\theta\partial_{\phi}- \cot\theta\partial_{\psi}),~~~~K^{\text{R}}_3=\partial_{\psi},\\[2mm]
		K^{\text{L}}_1+i K^{\text{L}}_2 &= e^{-i\phi}(i\partial_{\theta} + \cot\theta\partial_{\phi}- \csc\theta\partial_{\psi}),~~~~K^{\text{L}}_3=\partial_{\phi}.
\end{align}
Viewed as one forms these obey
\beq
dK_i^{\text{L}} + \frac{1}{2}\epsilon_{ijk}K_j^{\text{L}}\wedge K_k^{\text{L}}= dK_i^{\text{R}} - \frac{1}{2}\epsilon_{ijk}K_j^{\text{R}}\wedge K_k^{\text{R}}=0,
\eeq
and are given by
\begin{align}
K^{\text{R}}_1+i K^{\text{R}}_2&=e^{i \psi}(i d\theta + \sin\theta d\phi),~~~K^{\text{R}}_3=d\psi+\cos\theta d\phi,\\[2mm]
K^{\text{L}}_1+i K^{\text{L}}_2&=e^{-i \phi}(i d\theta -\sin\theta d\psi),~~~K^{\text{L}}_3=d\phi+\cos\theta d\psi.
\end{align}
These expressions can be repackaged as
\begin{align}
K_{i }^L= -i \text{Tr}\big[\sigma_i dg g^{-1}\big],~~~K_{i}^ R= -i \text{Tr}\big[\sigma_i g^{-1} dg \big],~~~ g=  e^{\frac{i}{2} \sigma_3\phi}e^{\frac{i}{2} \sigma_2\theta}e^{\frac{i}{2} \sigma_3\psi},
\end{align}
which makes clear that the  SU(2)$_{\text{L}/\text{R}}$ charged forms are SU(2)$_{\text{R}/\text{L}}$ invariant.

With these definitions it is not hard to confirm that \eqref{eq:spinordoublet} obeys \eqref{eq:S2lie}, which is the fundamental relations one needs to construct the SO(4) spinors.
\section{Proof that $\mathcal{N}=1$ implies $\mathcal{N}=4$ for SO(4) spinors}\label{sec:proof}
In section \ref{sec:one} it is claimed that  if an $\mathcal{N}=1$ sub-sector of the 7 dimensional SO(4) spinors of \eqref{eq:SO4spinors} solves the supersymmetry conditions, the entire $\mathcal{N}=4$ spinor also does. In this appendix we prove this claim using an argument based on studying (in abstract) the conditions  \eqref{eq10dspinors} must satisfy to have vanishing gravition and dilaton variation - see for example (2.10)-(2.14) of \cite{Kelekci:2014ima} for their explicit expressions in type IIA and IIB supergravity.
 
Since the solutions we consider respect the isometries of AdS$_3$ the 10 supersymmetry conditions that \eqref{eq10dspinors} must satisfy are implied by 7 dimensional conditions on $\chi^I_{1,2}$ only. For clarity let us assume that $\chi^I_{1}=\chi^I_{2}=\chi^I$ - the proof of the general case is in essence the same, just more cumbersome to describe. Schematically the independent 7 dimensional spinoral conditions for $\mathcal{N}=4$ then take one of two forms\footnote{The 10 dimensional dilatino and gravitino conditions along AdS$_3$ give rise to 7 dimensional conditions of the from \eqref{eq:7dsketch1}, while the gravitino condition along the internal space give rise to a condition of the from \eqref{eq:7dsketch1}.}
\begin{subequations}
\begin{align}
&\Delta_1\chi^I=0,\label{eq:7dsketch1}\\[2mm]
&(\nabla_{\mu}+ \Delta_2 \gamma_{\mu})\chi^I=0\label{eq:7dsketch2},
\end{align}
\end{subequations}
for $I=1,..,4$, where $\Delta_{1,2}$ are 8$\times$8 matrices containing combinations of the physical fields (dilaton, metric and RR and NS fluxes) and their derivatives contracted with the 7 dimensional gamma-matrices $\gamma_{\mu}$ and  where $\nabla_{\mu}$ is the covariant derivative. If one now decomposes
\beq
\chi^I= (\chi^i,~\chi^4),~~~ i=1,2,3
\eeq
it  follows from \eqref{eq:SO4action} that 
\beq
\chi^i=\frac{1}{2}\mathcal{L}_{K^1_i}\chi^4=\frac{1}{2}\mathcal{L}_{K^2_i}\chi^4 =\mathcal{L}_{K^+_i}\chi^4.
\eeq
It should then be clear that if one assumes that just the 4th component of \eqref{eq:7dsketch1}-\eqref{eq:7dsketch2} is solved then one generates the other 3 automatically by acting with the spinorial derivative along the Killing vectors provided
\beq\label{eq:Neq1impNeq4cond}
[\mathcal{L}_{K^{1,2}_i},~\Delta_1]\chi^4=[\mathcal{L}_{K^{1,2}_i},~(\nabla_{\mu}+ \Delta_2 \gamma_{\mu})]\chi^4=0.
\eeq
The first condition holds trivially whenever $\Delta_1$ is an SO(4) singlet, which is true whenever one imposes this condition on the physical fields as we are. The second condition is a little trickier - while $\Delta_2$ commutes with the spinoral Lie derivative for the same reason as $\Delta_1$ the individual $\mu$ indexed terms in \eqref{eq:7dsketch2} do not in general commute by themselves. One can proceed by rewriting the 4th component of  \eqref{eq:7dsketch2} in the equivalent form
\begin{subequations}\label{eq:altgrav}
\begin{align}
&(\nabla_{r}+ \Delta_2 \gamma_{r})\chi^4=0,\label{eq:7dsketcha}\\[2mm]
&(\nabla_{K^{1}_i}+ \Delta_2 \slashed{K}^1_i)\chi^4=0\label{eq:7dsketchb},\\[2mm]
&(\nabla_{K^{2}_i}+ \Delta_2 \slashed{K}^2_i)\chi^4=0\label{eq:7dsketchc},
\end{align}
\end{subequations}
where $\nabla_{K}= K^{\mu}\nabla_{\mu}$ and $\slashed{K}= K^{\mu}\gamma_{\mu}$.
The form of the metric \eqref{eq:M7} ensures that one can always choose a frame where $\nabla_{r}= \partial_r$ so that \eqref{eq:7dsketcha} commutes with $\mathcal{L}_{K^{1,2}_i}$ trivially.  The proofs that \eqref{eq:7dsketchb} and \eqref{eq:7dsketchc} commute are essentially the same so we focus on the former: The 7 dimensional derivative term in \eqref{eq:7dsketchb} decompose as
\beq
\nabla_{K^{1}_i} =K^{1\mu}_i\nabla^{S^3_1}_{\mu}- \frac{1}{2} \slashed{\partial C_1} \slashed{K}^{1}_i 
\eeq 
where  $\nabla^{S^3_1}_{\mu}$ obeys the equation \eqref{eq:S3KSE} when it acts on the components of the $\xi^a_1$ factor of $\chi^4$ - as such we can bring \eqref{eq:7dsketchb} into the form
\beq\label{eq:condth}
\Delta_3 \slashed{K}^1_i\chi^4=0,
\eeq
for $\Delta_3$ a new 8$\times$8 matrix that is an SO(4) singlet. This is close to the required result, which now follows if one can commute $\mathcal{L}_{K^1_j}$ past $\slashed{K}^1_i$. To achive this, one more piece of information about the spinors on $S^3$ is required, namely that
\beq\label{eq:condthing}
\slashed{K}^{1,2}_i \xi^a_{1,2}= \frac{1}{2}(\sigma_i)^a_{~b}\xi^a_{1,2},
\eeq
which is easy to verify, for instance one can read it off from \eqref{eq:matrixbispinorS3}.
Given \eqref{eq:condthing} and \eqref{eq:S2lie} it is quick to establish that
\beq\label{eq:comutingid}
\mathcal{L}_{K^{1,2}_i} \slashed{K}^{1,2}_j \xi_{1,2}^a= \slashed{K}^{1,2}_i\mathcal{L}_{K^{1,2}_j }\xi^a_{1,2}.
\eeq
Thus if one acts on \eqref{eq:condth}  with $\mathcal{L}_{K^{1}_j}$ one finds 
\beq
\mathcal{L}_{K^{1}_j}(\Delta_3 \slashed{K}^1_i\chi^4)=\Delta_3\slashed{K}^1_j \mathcal{L}_{K^{1}_i}\chi^4 =2\Delta_3\slashed{K}^1_j\chi^i=0,
\eeq
where the first equality follows from \eqref{eq:comutingid} and because $\Delta_3$ is an SO(4) singlet. Repeating the same steps for  \eqref{eq:7dsketchc} one establishes that
\begin{subequations}
\begin{align}
&(\nabla_{r}+ \Delta_2 \gamma_{r})\chi^i=0,\label{eq:7dsketcha}\\[2mm]
&(\nabla_{K^{1}_j}+ \Delta_2 \slashed{K}^1_i)\chi^i=0\label{eq:7dsketchb},\\[2mm]
&(\nabla_{K^{2}_j}+ \Delta_2 \slashed{K}^2_i)\chi^i=0\label{eq:7dsketchc},
\end{align}
\end{subequations}
are implied by \eqref{eq:altgrav}, from which it follows that the 4th component of \eqref{eq:7dsketch2} implies the other 3, which completes the proof. 

One can extend this argument to the case where $\chi_1^I \neq  \chi_2^I$ without difficulty, so solving the supersymmetry constraints for an $\mathcal{N}=1$ sub-sector of \eqref{eq:SO4spinors} implies that all components of $\chi^I_{1,2}$ also solve these constraints, provided the physical fields are SO(4) singlets. Thus solving an $\mathcal{N}=1$ sub-sector is sufficient to know that $\mathcal{N}=4$ supersymmetry  and an SO(4) R-symmetry is preserved by all solutions consistent with $\chi^I_{1,2}$.

\end{document}